\documentclass[3p,times,twocolumn]{elsarticle}

\usepackage{ecrc}

\volume{00}

\firstpage{1}

\journalname{Nuclear Physics B Proceedings Supplement}

\runauth{Kolja Kauder \emph{for the STAR Collaboration}}

\jid{nppp}

\jnltitlelogo{Nuclear Physics B Proceedings Supplement}

\usepackage{amssymb}
\usepackage{amsthm}

\usepackage{amsmath}

\usepackage[usenames,dvipsnames]{color}

\usepackage[percent]{overpic}

\usepackage{lineno}

\usepackage[figuresright]{rotating}

\newcommand{\pT}{\ensuremath{p_{T}}}

\newcommand{\sqrtsn}{\ensuremath{\sqrt{s_{NN}}}}
\newcommand{\AuAu}{Au+Au}
\newcommand{\PbPb}{Pb+Pb}
\newcommand{\pp}{p+p}
\begin{document}

\begin{frontmatter}



\dochead{}

\title{Di-Jet Imbalance Measurements in Central Au+Au Collisions at $\sqrt{s_{NN}}=200$~GeV from STAR}


\author{Kolja Kauder \emph{for the STAR Collaboration}}

\address{Wayne State University}

\begin{abstract}
The STAR collaboration reports the first measurements of the transverse momentum asymmetry $A_J$ 
of di-jet pairs in central gold-gold collisions
and minimum bias proton-proton collisions at $\sqrt{s_{NN}}=200$~GeV at RHIC.
We focus on anti-$k_T$ di-jets with
a leading jet $p_T>20$~GeV/$c$ and a subleading jet $p_T>10$~GeV/$c$, with a
constituent cut of 2~GeV/$c$, which reduces the effect of the underlying
heavy-ion background.
We examine the evolution of $A_J$ while reclustering these same di-jets with a lower constituent cut of 200 MeV/$c$.

For the low \pT\ constituent cut with a resolution parameter of $R=0.4$, the balance
between the di-jets is restored to the level of p+p collisions which
indicates the \emph{lost energy} observed for di-jets with a constituent cut of
 $p_T ^{\text{Cut}}>2$~GeV/$c$ is recovered.
Further variations of $R$ and the constituent \pT-cutoff indicate that
the lost energy is redistributed in the form of soft particles, accompanied by a broadening of the jet structure.
\end{abstract}

\begin{keyword}
di-jet imbalance


\end{keyword}

\end{frontmatter}

%\linenumbers


%%%%%%%%%%%%%%%%%%%%%%%%%%%%%%%%%%%%%%%%%%%%%%%%%%%%%%%
\section{Introduction}
\label{sec:intro}

The properties of the quark gluon plasma (QGP) formed in
high-energy nucleus-nucleus collisions can be studied using highly
energetic (hard) partons resulting from hard scatterings in the initial
stages of the collision, which fragment and hadronize into collimated cones of
particles known as jets.
Jets in proton-proton (\pp) collisions at RHIC are well-described by perturbative quantum chromodynamics (pQCD)
and can therefore be used as a reference to study medium-induced jet modifications~\cite{star_jet}.
Advances in jet-finding techniques~\cite{fastjet}, and the proliferation of high-\pT\ jets at the higher energies accessible
at the Large Hadron Collider (LHC) with a center-of-mass energy per nucleon pair of \sqrtsn= 2.76~TeV, have made it possible
to study fully reconstructed jets in heavy-ion collisions.
Inclusive jet spectra in the most central \PbPb\ collisions are found to be suppressed 
when compared to scaled \pp\ or peripheral \PbPb\ measurements at the same collision energy.
The observed suppression is visible for jets with $\pT\sim40-210$~GeV~\cite{Adam:2015ewa,cmsjet,Aad:2012vca}, %~\cite{Adam:2015ewa, CMSJetRAA, AAad:2012vca} 
and for jets reconstructed with a resolution parameter as large as R=0.5.

Among the first results of differential jet quenching studies at the LHC was the observation of a striking energy imbalance for highly energetic
back-to-back di-jet pairs~\cite{atlasjet,cmsjet}.
This imbalance is defined as $A_{J}\equiv(p_{T,\text{lead}}-p_{T,\text{sublead}})/(p_{T,\text{lead}}+p_{T,\text{sublead}})$, 
where $p_{T,\text{lead}}$ and $p_{T,\text{sublead}}$ are the transverse momenta of the leading and sub-leading jet, respectively,
and the di-jets are required to be approximately back-to-back.
This observable shows a reduced sensitivity to detector effects and the underlying heavy-ion event
with respect to inclusive measurements and other di-jet observables.
Furthermore, when di-jets with a large energy imbalance were selected at the LHC,
it was observed that most of the \emph{lost energy} of these  jets seems to be recovered at low momentum and at large angles
with respect to the di-jet axis (more than $0.8$~sr away)~\cite{Chatrchyan:2012nia, atlasjetshape}.
By contrast, at RHIC energies, recent measurements based on correlations of hadrons with leading reconstructed jets or non-decay (direct) photons
\cite{jhcorr, phenix_gdir_jet} indicate that the lost energy remains much closer to the jet axis.
We present the first di-jet imbalance measurement in central \AuAu\ collisions at RHIC,
thus allowing a more direct comparison to jet quenching measurements at the LHC.

%%%%%%%%%%%%%%%%%%%%%%%%%%%%%%%%%%%%%%%%%%%%%%%%%%%%%%%
\section{Analysis Details}
\label{sec:details}
The data used in this analysis was collected by the STAR detector in \pp\ and \AuAu\ collisions at $\sqrt{s_{NN}}$ = 200~GeV in 2006 and 2007, respectively.
Charged tracks are reconstructed with the Time Projection Chamber (TPC)~\cite{Anderson:2003ur}, 
and the transverse energy ($E_T$) of neutral hadrons is measured in the Barrel Electromagnetic Calorimeter (BEMC) ~\cite{Beddo:2002zx}.
To avoid double counting the energy of charged hadrons that leave
energy within the BEMC, we employ a so-called full hadronic correction scheme,
in which the transverse momentum of any charged track that extrapolates to a tower is subtracted from the transverse energy of that tower.
Tower energies are not allowed to become negative via this correction. 

Events were selected by an online high tower (HT) trigger, which required $E_T > 5.4$~GeV in at least one BEMC tower.
In \AuAu\ collisions, only the 0-13\% most central events are analyzed,
where event centrality is a measure of the overlap of the colliding nuclei, determined by the uncorrected charged particle multiplicity
in the TPC within $|\eta| < 0.5$. Events are restricted to have a primary vertex position along the beam axis of $|v_z| < 30$ cm.
Tracks are required to have at least 20 points measured in the TPC (out of a maximum of 45),
a distance of closest approach (DCA) to the collision vertex of less than 1 cm,
and pseudorapidity within $|\eta| < 1$. 

Jets are reconstructed from charged tracks measured in the TPC and neutral particles in the BEMC,
using the anti-$k_{T}$ algorithm from the FastJet package~\cite{fastjet} with resolution parameters $R = 0.4$ and $0.2$.
The reconstructed jet axes are required to be within $|\eta| < 1-R$ to avoid edge effects.
In this analysis, the initial definition of the di-jet pair considers only tracks and towers with $p_{T}^{\text{Cut}} > 2$ GeV/$c$~in
the jet reconstruction 
to minimize the effects of background fluctuations and combinatorial jets, as well as to make it unnecessary to
subtract the average background energy from the jets.
For a constituent $p_T$ cut of $p_{T}^{\text{Cut}} > 0.2$ GeV/$c$ the event-by-event average background energy is subtracted utilizing the density $\rho$, determined
as the median of $p_T^{\text{jet,rec}} / A^{\text{jet}}$ of all but the two leading jets,
using the $k_T$ algorithm with the same resolution parameter $R$ as in the nominal jet reconstruction.
The area $A^{\text{jet}}$ of jets is found within the FastJet package using active ghost particles~\cite{fastjet}.
For this part of the analysis, the corrected jet $p_T = p_T^{\text{jet,rec}}  - \rho A^{\text{jet}}$ will be used.

The di-jet imbalance $|A_J|$ is calculated for the highest and second-highest $p_T$ (leading and sub-leading) jets in \AuAu\ HT events,
provided they are approximately back-to-back ($|\phi_\text{lead} - \phi_\text{sublead} - \pi | < 0.4$)
and fulfill $p_{T,\text{lead}}>20$ GeV/$c$ and $p_{T,\text{sublead}}>10$ GeV/$c$.
%\textcolor{Red} {A bit ambiguous, since we only require this for pT$>$2, clarify!}
%
In order to make meaningful quantitative comparisons between the di-jet imbalance measured in \AuAu\ to that in \pp,
it is necessary to compare jets which have similar initial parton energies in the two collision systems,
and to take the remaining effect of background fluctuations into account.  
Therefore a di-jet imbalance reference data set is constructed in this analysis by embedding \pp\ HT events into
minimum bias \AuAu\ events in the same centrality class (\pp\ HT $\otimes$ \AuAu\ MB). 
The jet energies are not corrected back to the original parton energies.
During embedding, we account for the differences between Au+Au and p+p 
in tracking efficiency in the TPC ($90\% \pm 7\%$),
relative tower efficiency ($98\% \pm 2\%$, negligible),
and the relative tower energy scale ($100\% \pm 2\%$).
Systematic uncertainty on $|A_J|$ was assessed in this process by varying the relative efficiency and tower scale within their uncertainties
and is shown in the \pp\ HT $\otimes$ \AuAu\ MB embedding reference  as colored shaded boxes.

%%%%%%%%%%%%%%%%%%%%%%%%%%%%%%%%%%%%%%%%%%%%%%%%%%%%%%%
\section{Results}
\label{sec:results}
 
\begin{figure}[t]
  \begin{overpic}[width=0.495\textwidth]{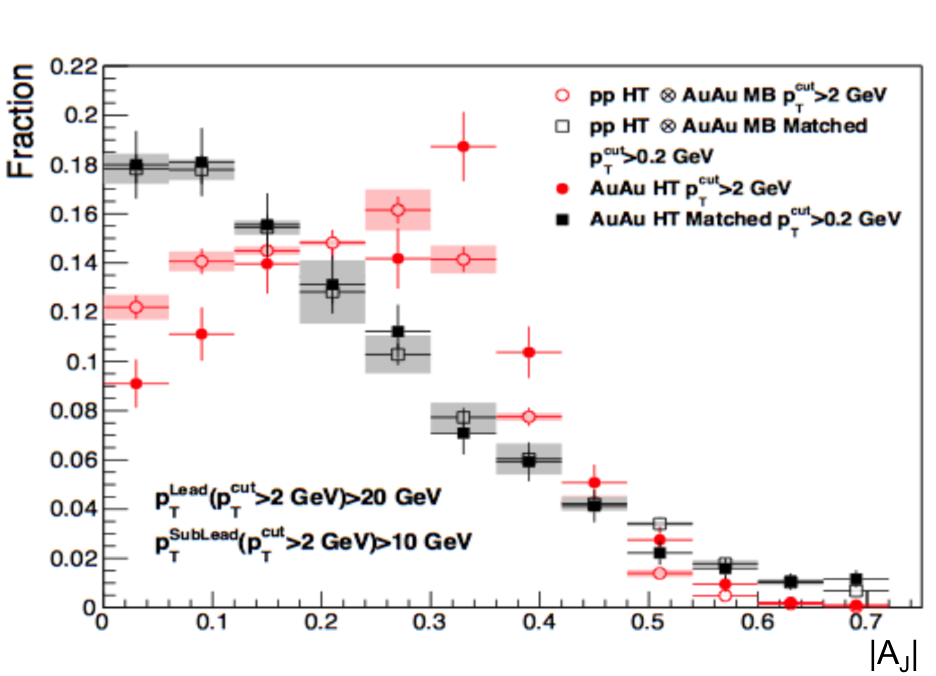}
  \put(65,27){\includegraphics [width=0.08\textwidth]{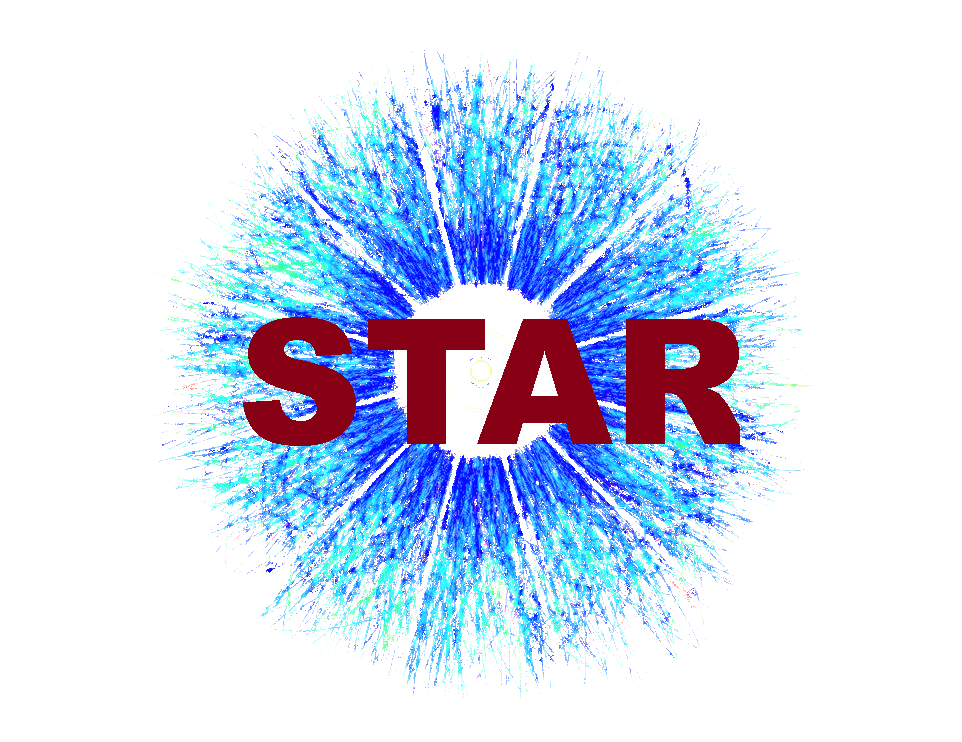}}
  \put(76,30){\textsf{\footnotesize Preliminary}}
  \end{overpic}
\caption{\label{fig:aj04}(Color online.)  
    Normalized $|A_J|$ distributions for \AuAu\ HT data (filled symbols) and p+p HT $\otimes$ \AuAu\ MB
    (open symbols). The red data points are for jets found using only constituents with $p_T^{\text{Cut}}>2$~GeV/$c$
    and the black ones for matched jets found using constituents with $p_T^{\text{Cut}}>0.2$~GeV/$c$.
    In all cases R=0.4.}
\end{figure}

In Fig.~\ref{fig:aj04} the $|A_J|$ distribution from central \AuAu\ collisions for anti-$k_T$ jets with $R=0.4$ (solid red markers)
is compared to the \pp\ HT embedding reference (\pp\ HT $\otimes$ \AuAu\ MB, open red markers)
for a jet constituent $p_T$-cut of $p_{T}^{\text{Cut}} > 2$ GeV/$c$.
Di-jets in central \AuAu\ collisions are significantly more imbalanced than the corresponding \pp\ di-jets.
To quantify this difference, the p-value for the hypothesis that the two histograms represent identical distributions
was calculated via a $\chi^2$-test~\cite{2006physics:5123G} including only the statistical uncertainties.
We find a p-value below $10^{-4}$, affirming the hypothesis that the  \AuAu\  and \pp\ HT $\otimes$ \AuAu\  data are
not drawn from the same parent  $A_J$ distributions.

In order to assess if the di-jet imbalance can be restored by including the jet constituents below 2~GeV/$c$ in transverse momentum,
the jet-finder was run again on the same events but lowering the constituent $p_T$-cut to $p_{T}^{\text{Cut}} > 0.2$ GeV/$c$. 
The absolute di-jet imbalance $|A_J|$ was then recalculated for low constituent \pT\ di-jet pairs
that could be geometrically matched to the initial di-jet pairs reconstructed with $p_{T}^{\text{Cut}} >2$~GeV/$c$.
Di-jet pairs were considered matched if their axes were within $\Delta R = \sqrt{\Delta \phi^{2} + \Delta \eta^{2} }<R$.
The geometric match is the only requirement, no \pT\ or other constraints were enforced.
The absolute di-jet imbalance $|A_J|$ for these matched low constituent \pT\ di-jet pairs was recalculated,
using background-corrected jet \pT.
The reference \pp\ HT $\otimes$ \AuAu\ MB embedding distribution was also recalculated to account for the effect
of the increased background fluctuations resulting from this low constituent $p_T$-cut.

In Fig.~\ref{fig:aj04} the matched di-jet imbalance measured for a low constituent $p_T$-cut in central \AuAu\ collisions (solid black markers)
is compared to the \pp\ HT $\otimes$ \AuAu\ MB embedding reference (open black markers).
The $|A_J|$ distribution in \AuAu\ is comparable to the \pp\ data within uncertainties (the p-value between these two distributions is 0.8),
suggesting that the jet energy balance can be restored to the level of p+p in central \AuAu\ HT events 
including low $p_T$ constituents and with an anti-$k_T$ jet of resolution parameter $R=0.4$.

%%%%%%%%%%%%%%%%%%%%%%%%%%%%%%%%%%%%%%%%%%%%%%%

\begin{figure}[t]
  \begin{overpic}[width=0.495\textwidth]{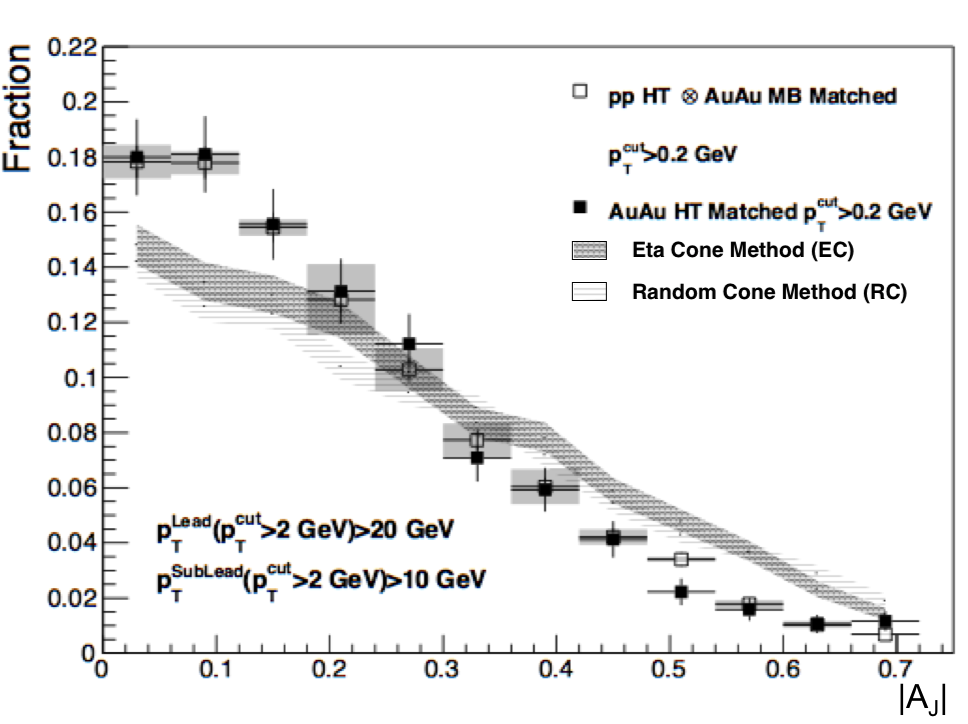}
  \put(65,27){\includegraphics [width=0.08\textwidth]{STAR-logo-base-red}}
  \put(76,30){\textsf{\footnotesize Preliminary}}
  \end{overpic}
\caption{\label{fig:aj04null}(Color online.) 
  Normalized $|A_J|$ distributions for \AuAu\ data (filled symbols) and \pp\ HT $\otimes$ \AuAu\ MB  (open symbols)
  for di-jets found using constituents with $p_{T}^{\text{Cut}}>0.2$~GeV/$c$ and matched to di-jets found with $p_{T}^{\text{Cut}}>2$~GeV/$c$. R=0.4.
  The bands indicate the $|A_J|$ distributions calculated assuming the RC and EC null hypotheses respectively; see the text for details.}
\end{figure}

The increase in background fluctuations for the low constituent $p_T$-cut 
could lead to an artificial di-jet energy balance unrelated to potential modifications of the jet fragmentation.
To estimate the magnitude of this effect, we employed the two different \emph{null hypothesis} procedures.
First, we embedded the \AuAu\ HT di-jets, reconstructed with a constituent $p_T$-cut $p_{T}^{\text{Cut}} > 2$ GeV/$c$,
(closed red markers in Fig.~\ref{fig:aj04}) into \AuAu\ MB events with a low constituent $p_T$-cut and re-calculated $|A_J|$.
This procedure explicitly disallows for any balance restoration via correlated signal jet constituents since the jet is embedded
 into a different random event.
We refer to this as the Random Cone (RC) technique.
In the second method, in order to account for potential non-jet correlations within the event,
we embed the same di-jet pairs as in the RC method into a different \AuAu\ HT event
with a found di-jet pair, at the same azimuth position but randomly offset in pseudorapidity by at least $2\times R$.
This Eta Cone method (EC) preserves potential  background effects due to azimuthal correlations of  the underlying event with the di-jet
while also excluding any potential jet-like correlation below 2~GeV/$c$.
Both of these methods are shown in Fig.~\ref{fig:aj04null} and are compared to the measured matched $|A_J|$
distribution with a low constituent $p_T$-cut. We conclude that background fluctuations alone can not account for the observed rebalancing, 
confirming that the energy restored via low $p_T$ constituents is correlated with the jets' fragmentation. 

%%%%%%%%%%%%%%%%%%%%%%%%%%%%%%%%%%%%%%%%%%%%%%%

% \begin{figure}[t]
%   \begin{overpic}[width=0.495\textwidth]{Proc_Fig3}
%   \put(58,22){\includegraphics [width=0.08\textwidth]{STAR-logo-base-red}}
%   \put(69,25){\textsf{\footnotesize Preliminary}}
%   \end{overpic}
%   \caption{\label{fig:ajvariation} (Color online.)
%     Normalized $|A_J|$ distributions with $R=0.2$ and $p_T^{\text{Cut}}>0.2$~GeV/$c$ for the matched jets.
%     The original distribution for $R= 0.4$and $p_T^{\text{Cut}}>0.2$~GeV/$c$
%     is shown for comparison.
%   }
% \end{figure}

\begin{figure*}[t]
 \begin{overpic}[width=0.48\textwidth]{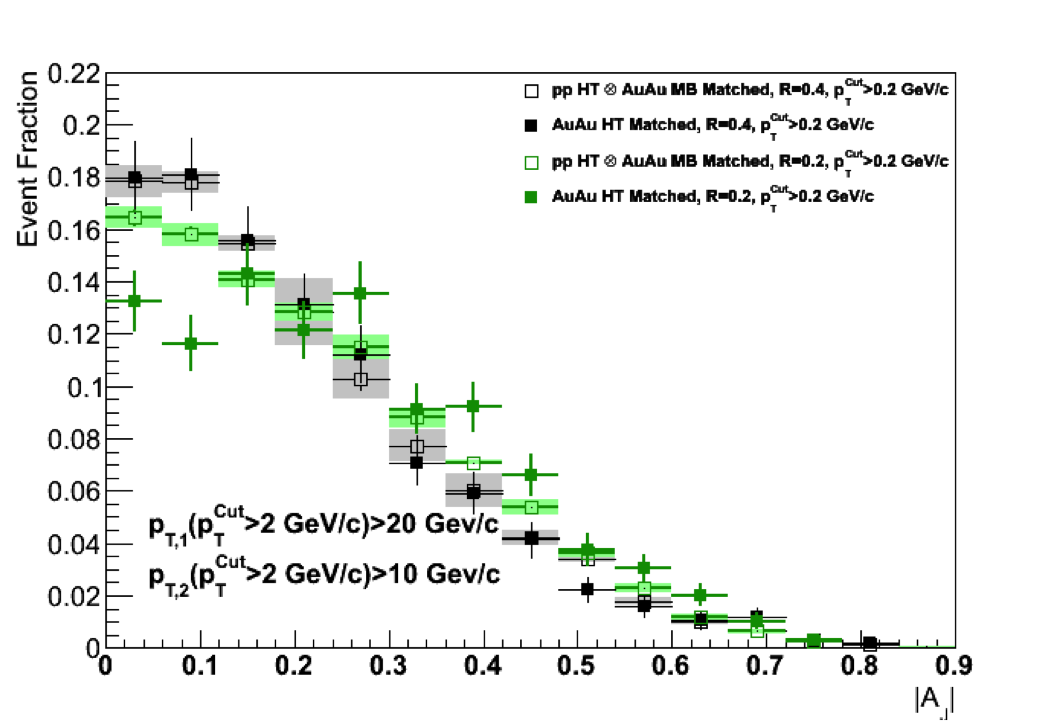}
  \put(58,27){\includegraphics [width=0.08\textwidth]{STAR-logo-base-red}}
  \put(69,30){\textsf{\footnotesize Preliminary}}
 \end{overpic}
 \begin{overpic}[width=0.48\textwidth]{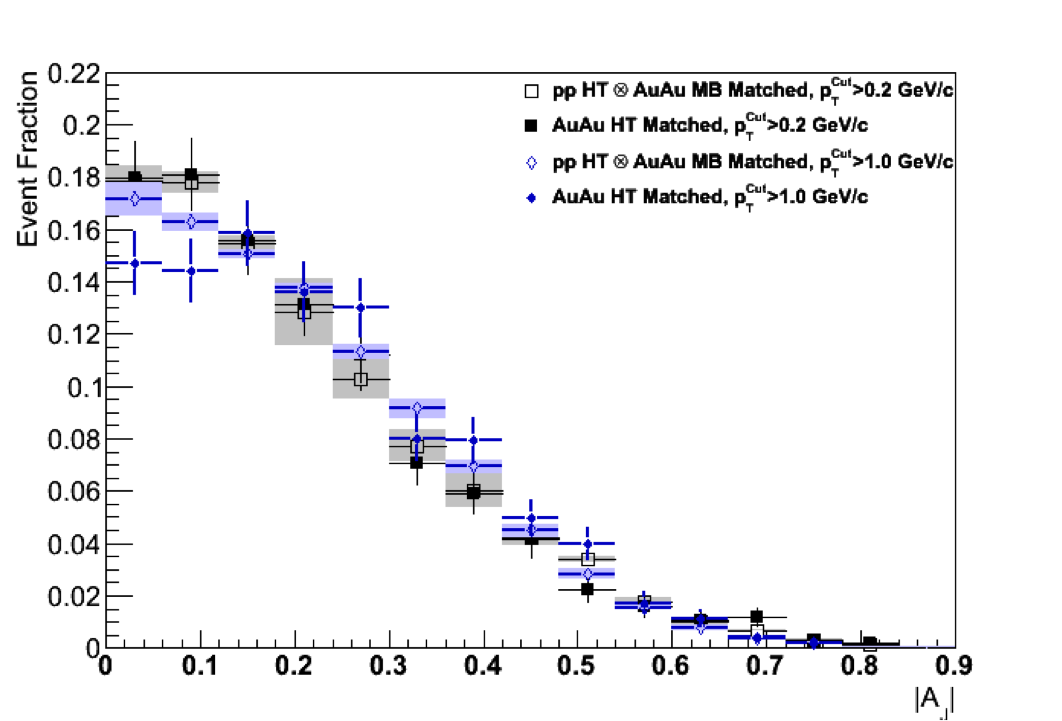}
  \put(58,27){\includegraphics [width=0.08\textwidth]{STAR-logo-base-red}}
  \put(69,30){\textsf{\footnotesize Preliminary}}
 \end{overpic}
 \caption{\label{fig:ajvariation} (Color online.)
   Normalized $|A_J|$ distributions with variations of jet-finding parameters for the matched jets.
   Left: Green squares show $R=0.2$ and $p_T^{\text{Cut}}>0.2$~GeV/$c$ for the matched jets.
   Right: Blue diamonds show $R=0.4$ and $p_T^{\text{Cut}}>1.0$~GeV/$c$. 
   The original distribution for $R= 0.4$ and $p_T^{\text{Cut}}>0.2$~GeV/$c$ is shown in black for comparison.
 }
\end{figure*}

%%%%%%%%%%%%%%%%%%%%%%%%%%%%%%%%%%%%%%%%%%%%%%%%%%%%%%%
%%%%%%%%%%%%%%%%%%%%%%%%%%%%%%%%%%%%%%%%%%%%%%%%%%%%%%%

For more differential insight into the energy loss structure, 
we repeat the second part of the measurement with two variations.
First, the $p_T^{\text{Cut}}$ for constituents is again lowered to $0.2$~GeV/$c$,
but this time the jet resolution parameter $R$ is changed to 0.2. 
Then we return to $R=0.4$ but now choose an intermediate 
$p_T^{\text{Cut}}$ of $1.0$~GeV/$c$.
Since the jet definition is changed, the embedded p+p reference is recalculated as well.
In both cases, the geometrical match to the same original hard-core di-jets
with $R=0.4$ and $p_{T}^{\text{Cut}} > 2$~GeV/$c$ is retained. 

As seen in Figure~\ref{fig:ajvariation},
narrowing the cone to $R=0.2$ leads to significant differences between 
central Au+Au and embedded p+p, even when including soft constituents down to 0.2~GeV/$c$ 
(p-value$\equiv2\times10^{-4}$).
This indicates broadening of the jet structure; jet energy, while contained within the original $R=0.4$ radius, 
is nevertheless transported away from the jet axis.
The difference between p+p and Au+Au for constituents above $1.0$~GeV/$c$ and $R=0.4$
indicates softening of the constituent spectrum in the medium,
albeit not quite as significant, with a p-value around 5\%.

%%%%%%%%%%%%%%%%%%%%%%%%%%%%%%%%%%%%%%%%%%%%%%%%%%%%%%%
\section{Conclusion}
\label{sec:conclusion}

In conclusion, these results indicate that even for a biased selection of di-jet pairs with hard cores
(constituents above 2 GeV/$c$), we observe some energy loss to the medium.
This lost energy seems to re-emerge as soft particles (below 2 GeV/$c$) 
within jets reconstructed with a resolution parameter of $R=0.4$, but with a broadened jet profile. % within $R=0.4$.
These findings indicate significant differences when compared to measurements at the LHC~\cite{Chatrchyan:2012nia,atlasjetshape},
in which the balance could only be restored when including constituents at large angles with respect to the di-jet axis.

Since our initial di-jet selection at RHIC contains a significant bias (constituents above 2 GeV/$c$),
it is conceivable that our kinematic selection leads to a pronounced surface bias of the 
di-jet creation point~\cite{Renk:2012hz,Renk:2012cx},
and subsequently a shorter but finite in-medium path-length with respect to the unbiased di-jet selection at LHC energies.
For future measurements with recently collected high-statistics data sets at STAR,
this scenario holds the tantalizing promise of
\emph{jet geometry engineering} of jet production points, path lengths and interaction probability within the colored medium
via variation of the jet-finding parameters and kinematic cuts.

%%%%%%%%%%%%%%%%%%%%%%%%%%%%%%%%%%%%%%%%%%%%%%%%%%%%%%%



%% References

%%\nocite{*}
\bibliographystyle{epj} %elsarticle-num}
\bibliography{Kauder_K}






% \end{thebibliography}

\end{document}